\RequirePackage{lineno}
\documentclass[aps,prd,showpacs,byrevtex,reprint, linenumber]{revtex4-1}

\usepackage{graphicx}
\usepackage{dcolumn}
\usepackage{bm}
\usepackage{rotating}
\usepackage{epstopdf}
\usepackage{booktabs}
\usepackage{color}
\usepackage{verbatim} 
\usepackage{multirow}
\usepackage[abs]{overpic}
\usepackage{amsmath}
\usepackage{amssymb}
\usepackage{xspace}
\usepackage[caption=false]{subfig}

\usepackage[colorlinks,
            linkcolor=blue,
            anchorcolor=blue,
            citecolor=blue]{hyperref}

\newcommand{\PreserveBackslash}[1]{\let\temp=\\#1\let\\=\temp}
\newcolumntype{C}[1]{>{\PreserveBackslash\centering}p{#1}}
\newcolumntype{R}[1]{>{\PreserveBackslash\raggedleft}p{#1}}
\newcolumntype{L}[1]{>{\PreserveBackslash\raggedright}p{#1}}

\newcommand{\EE}{e^+e^-}

\newcommand{\pip}{\pi^{+}}
\newcommand{\pim}{\pi^{-}}
\newcommand{\piz}{\pi^{0}}

\begin{document}
\graphicspath{{figure/}}
\DeclareGraphicsExtensions{.eps,.png,.ps}
\title{\boldmath Search for $\psi_0(4360)\rightarrow \eta\psi(2S)$ through the process $\EE \rightarrow \eta\eta\psi(2S)$}

\author{
\begin{small}
\begin{center}
M.~Ablikim$^{1}$\BESIIIorcid{0000-0002-3935-619X},
M.~N.~Achasov$^{4,d}$\BESIIIorcid{0000-0002-9400-8622},
P.~Adlarson$^{81}$\BESIIIorcid{0000-0001-6280-3851},
X.~C.~Ai$^{87}$\BESIIIorcid{0000-0003-3856-2415},
C.~S.~Akondi$^{31A,31B}$\BESIIIorcid{0000-0001-6303-5217},
R.~Aliberti$^{39}$\BESIIIorcid{0000-0003-3500-4012},
A.~Amoroso$^{80A,80C}$\BESIIIorcid{0000-0002-3095-8610},
Q.~An$^{77,64,\dagger}$,
Y.~H.~An$^{87}$\BESIIIorcid{0009-0008-3419-0849},
Y.~Bai$^{62}$\BESIIIorcid{0000-0001-6593-5665}
% ...
\\
\vspace{0.2cm}
(BESIII Collaboration)\\
\vspace{0.2cm} {\it
$^{1}$ Institute of High Energy Physics, Beijing 100049, People's Republic of China\\
% ...
$^{4}$ Budker Institute of Nuclear Physics SB RAS (BINP), Novosibirsk 630090, Russia\\
% ...
$^{31}$ INFN Sezione di Ferrara, (A)INFN Sezione di Ferrara, I-44122, Ferrara, Italy; (B)University of Ferrara, I-44122, Ferrara, Italy\\
% ...
$^{39}$ Johannes Gutenberg University of Mainz, Johann-Joachim-Becher-Weg 45, D-55099 Mainz, Germany\\
% ...
$^{62}$ Southeast University, Nanjing 211100, People's Republic of China\\
% ...
$^{64}$ State Key Laboratory of Particle Detection and Electronics, Beijing 100049, Hefei 230026, People's Republic of China\\
% ...
$^{77}$ University of Science and Technology of China, Hefei 230026, People's Republic of China\\
% ...
$^{80}$ University of Turin and INFN, (A)University of Turin, I-10125, Turin, Italy; (B)University of Eastern Piedmont, I-15121, Alessandria, Italy; (C)INFN, I-10125, Turin, Italy\\
$^{81}$ Uppsala University, Box 516, SE-75120 Uppsala, Sweden\\
% ...
$^{87}$ Zhengzhou University, Zhengzhou 450001, People's Republic of China\\
\vspace{0.2cm}
$^{\dagger}$ Deceased\\
% ...
$^{d}$ Also at the Novosibirsk State University, Novosibirsk, 630090, Russia\\
% ...
}
\end{center}
\vspace{0.4cm}    
\end{small}
}

\vspace{0.2cm}
\date{\today}

\begin{abstract}
Using data samples corresponding to an integrated luminosity of 0.9~fb$^{-1}$ collected with the BESIII detector operating at the BEPCII storage ring at the center-of-mass energies of 4.84, 4.92, and 4.95~GeV, we 
search for the exotic charmonium-like state with quantum numbers $J^{PC}=0^{--}$, $\psi_0(4360)$, in the process $\EE\rightarrow\eta\psi_0(4360)$ with $\psi_0(4360)\rightarrow\eta\psi(2S)$. No significant signal of the $\psi_0(4360)$ resonance state is observed. Upper limits on $\sigma(\EE\rightarrow\eta\psi_0(4360))\cdot\mathcal{B}(\psi_0(4360)\rightarrow
\eta\psi(2S))$ and $\sigma(\EE\rightarrow\eta\eta\psi(2S))$ at the 90\% confidence level are determined for each energy point.
\end{abstract}

\newcommand{\BESIIIorcid}[1]{\href{https://orcid.org/#1}{\hspace*{0.1em}\raisebox{-0.45ex}{\includegraphics[width=1em]{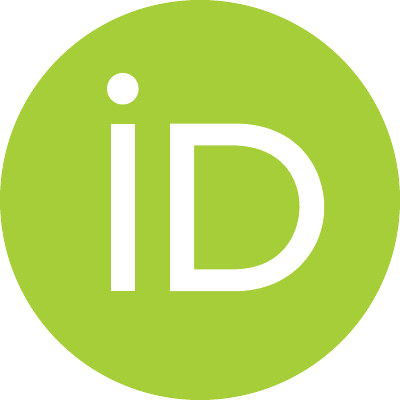}}}}

\maketitle
\section{INTRODUCTION}
Since the discovery of the $X(3872)$~\cite{Belle:2003_X3872}, many charmonium-like states, including the $Y(4260)$, $Y(4660)$, $T_{c\bar{c}1}(3900)$, and $T_{c\bar{c}}(4020)$ states, have been observed by the BaBar~\cite{BaBar:2005_Y4260}, Belle~\cite{Belle:2007umv}, and BESIII experiments~\cite{BESIII:2013ris,BESIII:2013ouc}.
They exhibit unconventional properties that are difficult to accommodate within the framework of the conventional quark model. These states are collectively referred to as $XYZ$ particles and regarded as promising candidates for exotic hadronic states. 
A wide range of theoretical interpretations have been proposed, including compact tetraquarks~\cite{Maiani:2014aja,Wang:2021qus}, hadronic molecules~\cite{Cleven:2013mka, Ding:2008gr, Wang:2013cya,Chen:2017abq}, hybrids~\cite{Zhu:2005hp, Close:2005iz, Kou:2005gt,Brambilla:2022hhi,Oncala:2017hop}, and hadro-charmonium states~\cite{Dubynskiy:2008mq, Li:2013ssa}.

Among the various $XYZ$ states observed in experiment, the vector charmonium-like states with $J^{PC} =1^{--}$ (known as $Y$ states) have attracted particular attention. 
In addition to the three well-established charmonium states $\psi(4040)$, $\psi(4160)$, and $\psi(4415)$, several other experimentally observed states such as $Y(4260)$, overpopulate the conventional charmonium spectrum above the open-charm threshold predicted by potential model~\cite{Brambilla:2019esw}.
The $Y(4260)$ state was discovered via the process $e^+e^- \to \gamma_{\text{ISR}}\pi^+\pi^- J/\psi$ by the BaBar experiment~\cite{BaBar:2005_Y4260} using the initial state radiation (ISR) process, and was subsequently confirmed by the CLEO~\cite{CLEO:2006ike} and Belle experiments~\cite{Belle:2007dxy}. 
The BESIII Collaboration performed a dedicated scan on the process $e^+e^- \to \pi^+\pi^- J/\psi$~\cite{BESIII:2016bnd}, and found two structures $Y(4230)$ (aka $\psi(4230)$) and $Y(4360)$ (aka $\psi(4360)$). 
Furthermore, structures similar to the $\psi(4230)$ and $\psi(4360)$ have been observed in various processes~\cite{BESIII:2022qal,BESIII:2020oph,PhysRevD.107.092005,BESIII:2022joj,PRD96032004,PRD104052012,PRL118092002,PRL114092003,PRD99091103,BESIII:2018iea,PhysRevLett.129.102003}. 
However, the parameters of these two resonances, as measured in different processes, are in tension with each other~\cite{Guo:2025ady}. 
Despite the diversity of proposed interpretations, including hybrid~\cite{Zhu:2005hp,Kou:2005gt,Close:2005iz}, hadro-charmonium~\cite{Dubynskiy:2008mq,Li:2013ssa}, conventional charmonium~\cite{Llanes-Estrada:2005qvr,Li:2013bca}, and bound systems~\cite{MartinezTorres:2009xb}, the hadronic molecular picture provides a compelling framework. Within this picture, the $\psi(4230)$, $\psi(4360)$, and $\psi(4415)$ can be understood as molecular configurations dominated by the $D\bar{D}_1$, $D\bar{D}_1(2420)$, and $D\bar{D}_2$ components, respectively~\cite{Wang:2013cya,Wang:2013kra,Ma:2014zva,Cleven:2015era,Hanhart:2019isz,Anwar:2021dmg}.

Recently, based on the molecular scenarios, an exotic charmonium-like
state with quantum numbers $J^{PC}=0^{--}$, denoted as
$\psi_0(4360)$~\cite{Ji:2022blw}, has been predicted as a bound system
of $D^*\bar D_1(2420)$.  The mass and width of this state are
predicted to be $(4366\pm18)$~MeV$/c^2$ and less than 10~MeV,
respectively.  It can be searched for in the process $e^+e^-\to \eta
\psi_0(4360)$ and distinguished from $1^{--}$ charmonium vector states by
the distinctly different angular distribution of the outgoing $\eta$
meson. Observation of a $0^{--}$ molecule would play a crucial role in
inferring the internal structure of the vector mesons in the mass
range between 4.2 and 4.5~GeV$/c^2$ and in understanding the nature of
other exotic hadronic states, such as
$Z_c$~\cite{BESIII:2013ris,BESIII:2013ouc,BESIII:2013qmu} and
$P_c$~\cite{LHCb:2015yax,LHCb:2019kea} states.

In this analysis, we search for the processes $\EE \rightarrow \eta\psi_0(4360)$ with $\psi_0(4360)\rightarrow \eta\psi(2S)$ 
using data samples collected at three center-of-mass (c.m.) energies ($\sqrt{s}$): 4.84, 4.92, and 4.95~GeV, corresponding to a total integrated luminosity of 0.9~fb$^{-1}$. 
This analysis is performed using a partial reconstruction method, where one $\eta$ is reconstructed via its decay to $\gamma\gamma$, while the other $\eta$ is left unreconstructed and treated inclusively. 
The $\psi(2S)$ candidate is reconstructed with $\psi(2S)\to\pip\pim J/\psi$, where $J/\psi$ is reconstructed by its leptonic decays, $J/\psi\rightarrow e^+e^-$ and $J/\psi\rightarrow\mu^+\mu^-$, referred to as the $ee$-mode and $\mu\mu$-mode, respectively.

\section{BESIII DETECTOR AND MONTE CARLO SIMULATION}

The BESIII detector~\cite{BESIIIDetector} records symmetric $e^+e^-$ collisions provided by the BEPCII storage ring~\cite{BEPCII} in the c.m. energy range from 1.84 to 4.95~GeV, with a peak luminosity of $1.1 \times 10^{33}\;\text{cm}^{-2}\text{s}^{-1}$ achieved at $\sqrt{s} = 3.773\;\text{GeV}$. BESIII has collected large data samples in this energy region~\cite{WhitePaper}. The cylindrical core of the BESIII detector covers 93\% of the full solid angle and consists of a helium-based multilayer drift chamber~(MDC), a time-of-flight system~(TOF), and a CsI(Tl) electromagnetic calorimeter~(EMC), which are all enclosed in a superconducting solenoidal magnet providing a 1.0~T magnetic field. The solenoid is supported by an octagonal flux-return yoke with resistive plate counter muon identification modules (MUC) interleaved with steel. 
The charged-particle momentum resolution at $1~{\rm GeV}/c$ is $0.5\%$, and the ${\rm d}E/{\rm d}x$ resolution is $6\%$ for electrons from Bhabha scattering. The EMC measures photon energies with a resolution of $2.5\%$ ($5\%$) at $1$~GeV in the barrel (end-cap) region. The time resolution in the plastic scintillator TOF barrel region is 68~ps, while that in the end-cap region was 110~ps. The end-cap TOF system was upgraded in 2015 using multigap resistive plate chamber technology, providing a time resolution of 60~ps, which benefits all data used in this analysis~\cite{tof1,tof2,tof3}.

Monte Carlo (MC) simulated samples produced with {\sc
geant4}-based~\cite{GEANT4} software, which includes the
geometric description of the BESIII detector and the detector
response, are used to determine detection efficiencies and to estimate
backgrounds.  The simulation models the beam-energy spread and ISR in
the $e^+e^-$ annihilations with the generator {\sc
kkmc}~\cite{KKMC}. The inclusive MC sample includes the production of
open-charm processes, the ISR production of vector charmonium(-like)
states, and the continuum processes incorporated in {\sc
kkmc}~\cite{KKMC}. All particle decays are modeled with {\sc
evtgen}~\cite{EVTGEN} using branching fractions either taken from the
Particle Data Group (PDG)~\cite{ParticleDataGroup:2024cfk}, when
available, or otherwise estimated with {\sc
lundcharm}~\cite{lundcharm1,lundcharm2}.  Final-state radiation from
charged final-state particles is incorporated using {\sc photos}~\cite{PHOTOS}.

Signal MC samples for the process $\EE \rightarrow \eta\psi_0(4360)$
with $\psi_0(4360)\rightarrow\eta\psi(2S)$ are generated 
considering a cascade two-body P-wave process. The
angular distribution of the cascade products follows a
$1-\cos^2\theta$ dependence at each c.m. energy. The mass and width of
$\psi_0(4360)$ are set to be 4366~MeV$/c^2$ and 10~MeV, respectively,
based on the prediction~\cite{Ji:2022blw}. MC samples for the process
$\EE \rightarrow \eta\eta\psi(2S)$ without a $\psi_0(4360)$ intermediate
state are generated using a phase-space (PHSP) model at each
c.m. energy.  In the MC simulation of the two signal
processes, both $\eta$ mesons are allowed to decay inclusively. The
decay $\psi(2S)\rightarrow\pip\pim J/\psi$ is generated using {\sc
evtgen}~\cite{EVTGEN} with the {\sc jpipi} model, and the subsequent
decays $J/\psi\rightarrow e^+e^-$ and $J/\psi\rightarrow \mu^+\mu^-$
are generated with the {\sc photos vll} model.

\section{EVENT SELECTION}

To improve the selection efficiency, a partial reconstruction method is used for the signal processes $\EE \rightarrow \eta\eta\psi(2S)$ and $\EE \rightarrow \eta\psi_0(4360)$ with $\psi_0(4360)\rightarrow\eta\psi(2S)$, where only one $\eta$ decaying to two photons is selected. 
Candidate events are required to have four charged tracks with zero net charge and at least two photon candidates.

 Charged tracks detected in the MDC are required to be within a polar angle ($\theta$) range of $|\rm{cos\theta}|<0.93$, where $\theta$ is defined with respect to the $z$-axis, which is the symmetry axis of the MDC. 
 For each track, the distance of closest approach to the interaction point (IP) must be less than 10\,cm along the $z$-axis, $|V_{z}|$, and less than 1\,cm in the transverse plane, $|V_{xy}|$.
 The pions and the leptons from the $\psi(2S)$ and $J/\psi$ decay are kinematically well separated.
 Tracks with momenta larger than 1.0~GeV/$c$ are assigned as leptons from the $J/\psi$ decay; otherwise, they are identified as pions from the $\psi(2S)$ decay. The leptons ($\ell^+\ell^-$, $\ell=e$ or $\mu$) from the $J/\psi$ decay with energy deposited in the EMC larger than 1.0~GeV are identified as electrons, while those with less than 0.4~GeV are identified as muons. 

Photon candidates are identified as isolated showers in the EMC. The deposited energy of each shower must be greater than 25~MeV in the barrel region ($|\cos \theta|< 0.80$) and greater than 50~MeV in the end-cap region ($0.86 <|\cos \theta|< 0.92$). 
To exclude showers that originate from
charged tracks, the opening angle subtended by the EMC shower and the position of the closest charged track at the EMC, measured from the IP, must be greater than 10 degrees. 
To suppress electronic noise and showers unrelated to the event, the difference between the EMC time and the event start time is required to be within [0, 700]\,ns.

In order to improve the mass resolution,
a one-constraint (1C) kinematic fit is performed by constraining the $\eta$ candidate reconstructed from a $\gamma\gamma$ pair to its nominal mass~\cite{ParticleDataGroup:2024cfk}. The other $\eta$ decays inclusively and is not reconstructed, but is instead treated as a missing particle. To further suppress background contributions, the invariant mass of the $\gamma\gamma$ pair before the 1C kinematic fit is required to be within the range of [499.5, 576.9] MeV/$c^2$. This mass window is optimized using the Punzi figure of merit (FOM)~\cite{Punzi:2003bu}, defined as $\epsilon/(1.5+\sqrt{B})$, where $\epsilon$ is the signal efficiency obtained from the signal MC samples and $B$ is the number of background events estimated from the dominant background process $e^+e^-\rightarrow\pi^0\pi^0\psi(2S)$ discussed in section.~IV. 
If there is more than one combination in an event, the one with minimum $\chi^{2}_{\text{1C}}$ value is retained.
For the $\mu\mu$-mode, we require at least one muon candidate to have a hit depth in the muon counter (MUC) greater than 30~cm to suppress the pion-to-muon misidentification.

For the two $\eta$ candidates in this analysis, the $\eta$
reconstructed from $\gamma\gamma$ is denoted as $\eta_1$, while
$\eta_2$ refers to the particle treated inclusively.  The mass of
$\eta_2$ is determined from the recoil against the selected
$\eta_1\psi(2S)$ system after the 1C kinematic fit, defined as
$M_{\eta_1\psi(2S)}^{\text {recoil
}}=\sqrt{\left(\sqrt{s}-E_{\eta_1\psi(2S)}\right)^2-\left|\vec{p}_{\eta_1\psi(2S)}\right|^2}$,
where $E_{\eta_1\psi(2S)}$ and $\vec{p}_{\eta_1\psi(2S)}$ are the
energy and momentum of the selected $\eta_1\psi(2S)$ candidate in the
c.m.~system, respectively. To reduce the impact of detector resolution
on $M(\ell^+\ell^-)$, the corrected mass of $\eta_2$ is obtained as
$M(\eta_2) = M_{\eta_1\psi(2S)}^{\text{recoil}} + M(\ell^+\ell^-) -
M(J/\psi)$,  where $M(J/\psi)$ denotes
the nominal masses of the $J/\psi$ from the
PDG~\cite{ParticleDataGroup:2024cfk}, and $M(\ell^+\ell^-)$ is the invariant
mass of the lepton pair. A similar correction is applied to the
invariant mass of $\pi^+\pi^-J/\psi$, which is defined as
$M(\pi^+\pi^-J/\psi) = M(\pi^+\pi^-\ell^+\ell^-) + M(\ell^+\ell^-) -
M(J/\psi)$, in order to reduce the detector resolution effect
originating from $M(\ell^+\ell^-)$.

The signal region for $\eta_2$ states is defined as 
[516.3, 579.5]~MeV/$c^2$, which is optimized using the Punzi FOM~\cite{Punzi:2003bu} in the same way as for the $\eta_1$ candidate. 
The invariant mass of the lepton pair $M$($\ell^+\ell^-$) is required to lie within [3040.0, 3140.8]~MeV/$c^2$ to identify the $J/\psi$. 
The $\psi(2S)$ signal region is determined from the signal MC as $\pm3\sigma$ (where $\sigma$ is the mass resolution) around the nominal mass value, corresponding to the range [3678.3, 3693.5]~MeV/$c^2$.

\section{BORN CROSS SECTION MEASUREMENT}

\subsection{\boldmath $\EE\rightarrow\eta\eta\psi(2S)$}

After applying the above requirements, the remaining background is mainly from $e^+e^-\rightarrow\piz\piz\psi(2S)$ events.
The number of expected background events in the $\psi(2S)$ signal region is estimated using
\begin{equation}
N_{\mathrm{bkg}} = 
\mathcal{L} \left(1+\delta^{\mathrm{bkg}}_{\mathrm{ISR}}\right)
\left(1+\delta_{\mathrm{VP}}\right)
\epsilon_{\mathrm{bkg}} \, 
\mathcal{B}_{\mathrm{bkg}} \, 
\sigma_{\mathrm{bkg}},
\label{func:cal-bkg}
\end{equation}
where ${\sigma_{\mathrm{bkg}}}$ is the measured Born cross section of $e^+e^-\rightarrow\piz\piz\psi(2S)$~\cite{ref:pi0pi0psip_lixuhong}, 
$\mathcal{L}$ is the integrated luminosity at each energy point, 
$\epsilon_{\mathrm{bkg}}$ represents the detection efficiency of the background process after applying the signal selection, and $\mathcal{B}_{\rm bkg}$ is the product of branching fraction of the intermediate states taken from the PDG~\cite{ParticleDataGroup:2024cfk}. The factor $(1+{\delta}^{\mathrm{bkg}}_{\mathrm{ISR}})$ accounts for the ISR correction quoted from Ref.~\cite{ref:pi0pi0psip_lixuhong}, and ${\left(1+\delta_{\mathrm{VP}}\right)}$ is the vacuum polarization (VP) correction factor~\cite{vacuum}, which depends only on the c.m. energy and is therefore universal for all processes. The numbers of background events are estimated to be 0.10, 0.05 and 0.02 at the three energy points. 

\begin{figure*}[htbp]
    \centering
     \includegraphics[width=0.43\textwidth]{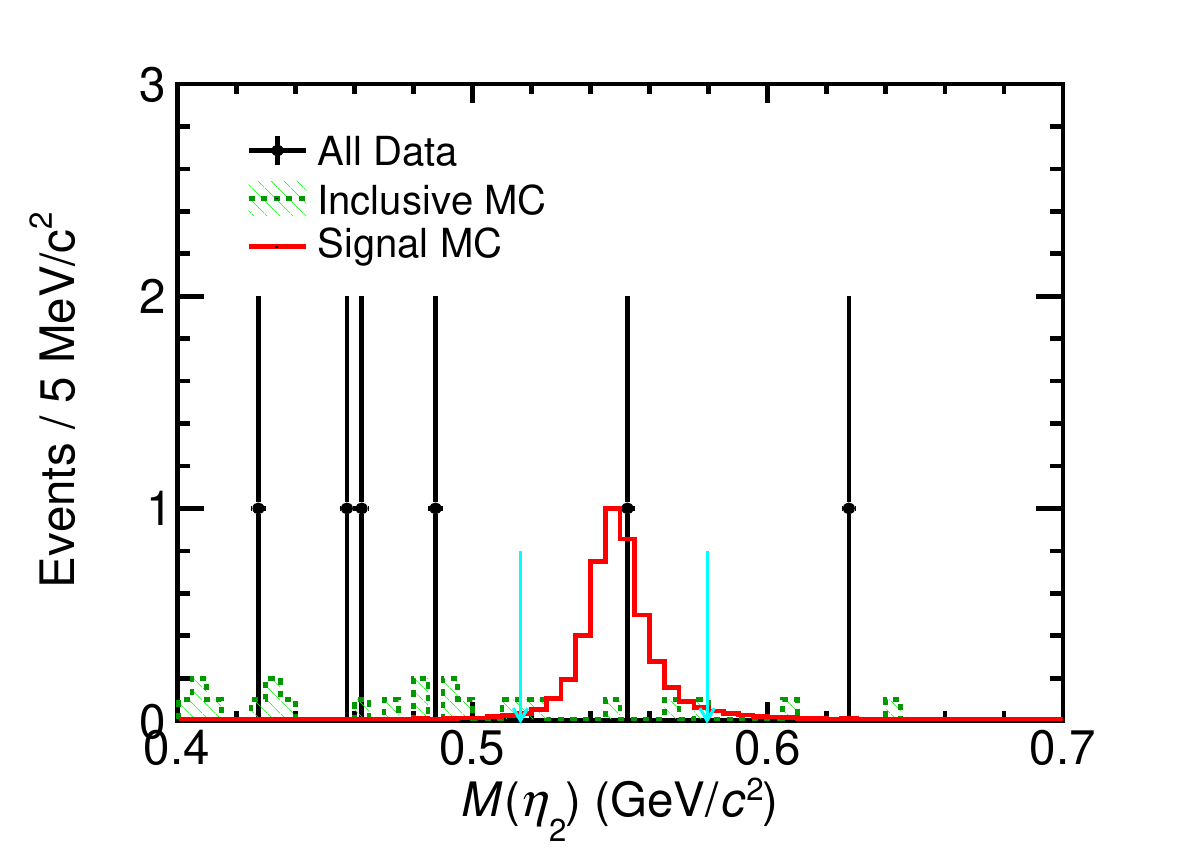}
    \includegraphics[width=0.43\textwidth]{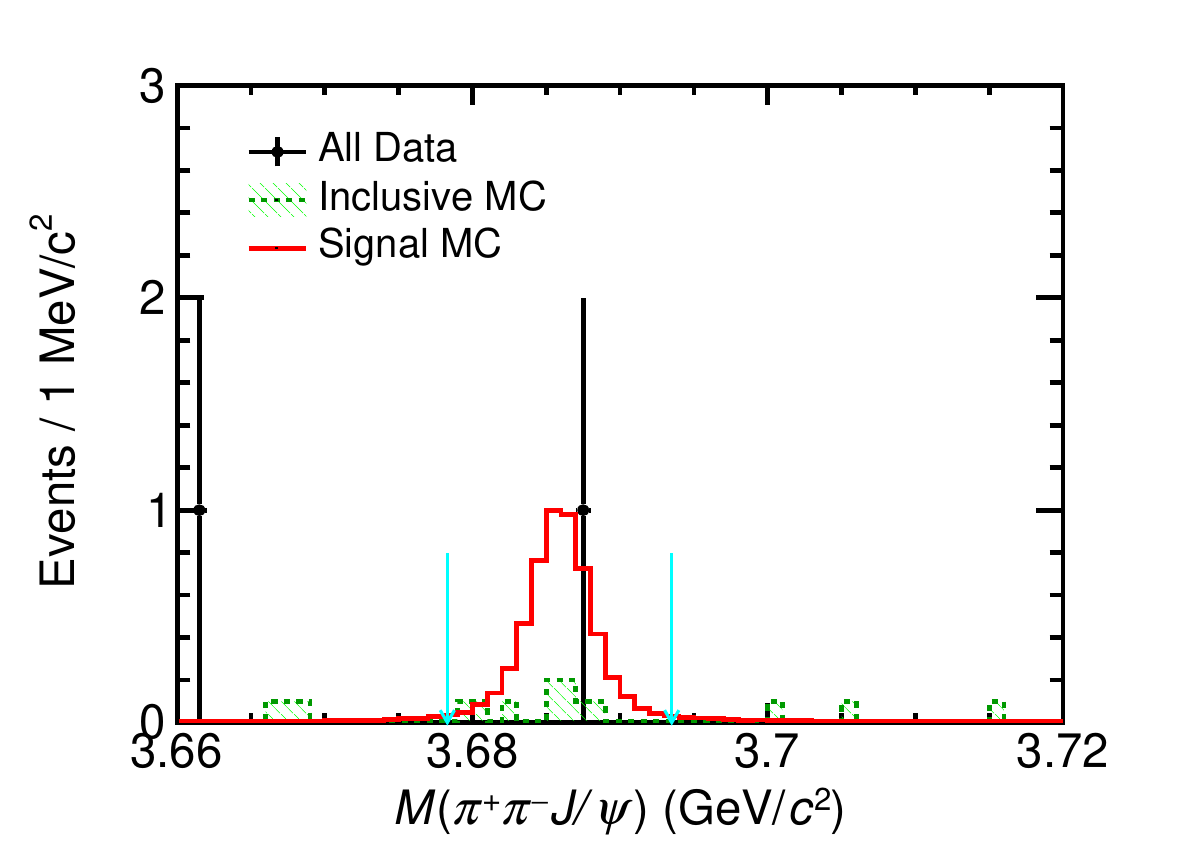}
 \caption{Distributions of $M(\eta_2)$ and $M(\pi^{+}\pi^{-}J/\psi)$ combined from all c.m. energies. The black dots with error bars denote the real data, the green dashed histograms denote the inclusive MC simulation, and the red histograms correspond to the signal MC simulation of $\EE\rightarrow\eta\eta\psi(2S)$. The blue vertical lines indicate the signal regions of $\eta_2$ and $\psi(2S)$.}
\label{fig:data_psi_3}
\end{figure*}

Figure~\ref{fig:data_psi_3} presents the distributions of the $\eta_2$
and $\pip\pim J/\psi$ invariant masses for the process
$\EE\rightarrow\eta\eta\psi(2S)$ for the combined datasets.  No
significant $\psi(2S)$ signal is observed.  The signal yield
($N_\text{sig}$) is obtained by subtracting the estimated background
from the observed number of events in the $\psi(2S)$ signal region,
i.e. $N_{\text{sig}}=N_{\text{obs}}-N_{\text{bkg}}$. The upper limits
on the numbers of signal events are calculated at the 90\% confidence
level~(CL) using a frequentist method. This method, described in
Ref.~\cite{trolke1}, incorporates an unbounded profile-likelihood
treatment for multiplicative systematic uncertainties and is
implemented via {\sc trolke}~\cite{trolke2} within the
{\sc root} framework~\cite{rootmanual}, where the numbers of signal
and background events follow Poisson statistics and the efficiencies
are modeled with Gaussian distributions.

The Born cross section of $\EE\rightarrow\eta\eta\psi(2S)$ is calculated with
\begin{equation}
\begin{aligned}
&\sigma(\EE\rightarrow\eta\eta\psi(2S))\\
&=\frac{N_{\text{sig}}}{\mathcal{L} \left(1+\delta^{\mathrm{sig}}_{\mathrm{ISR}}\right)\left(1+\delta_{\mathrm{VP}}\right) \mathcal{B}_{\psi(2S)}\left(\epsilon_{\mathrm{e}} \mathcal{B}_{\mathrm{e}}+\epsilon_{\mu} \mathcal{B}_{\mu}\right) }.
\end{aligned}
\label{eq:cal_cross}
\end{equation}
Here, $N_{\text{sig} }$ is the number of signal events, 
$(1+\delta^{\mathrm{sig}}_{\mathrm{ISR}})$ is the ISR correction factor obtained from a quantum electrodynamics calculation~\cite{Kuraev:1985hb} using the {\sc kkmc} generator~\cite{KKMC}, assuming that the cross section follows the lineshape of a three-body PHSP model, $\varepsilon_{e}$ and $\varepsilon_{\mu}$ are the detection efficiencies for the $ee-$mode and $\mu\mu-$mode from the signal MC simulation, respectively. 
$\mathcal{B}_{\psi(2S)}$ is the branching fraction for $\psi(2S) \to \pi^+\pi^- J/\psi$, and $\mathcal{B}_e$ and $\mathcal{B}_\mu$ are the branching fractions of the decays $J/\psi \to e^+e^-$ and $J/\psi \to \mu^+\mu^-$, respectively. 
The numerical results for each data sample are listed in Table~\ref{table:sum_result}.

\begin{table*}[htbp]
\centering
\caption{Numerical results for the process $e^+e^-\rightarrow\eta\eta\psi(2S)$ at each c.m. energy, where $\sqrt{s}$ is the c.m. energy, $\mathcal{L}$ is integrated luminosity, $N_{\mathrm{obs}}$ is the number of events in the $\psi(2S)$ signal region, 
$N_{\mathrm{sig}}$ is the calculated signal yield in the $\psi(2S)$ signal region, $N^{\mathrm{UL}}$ is the upper limit of signal yield at the 90\% CL, $(1+\delta^{\mathrm{sig}}_{\mathrm{ISR}})$ is the ISR correction factor, $(1+\delta_{\mathrm{VP}})$ is the vacuum polarization factor, $\epsilon$ is the detection efficiency, $\sigma^{\mathrm{Born}}$ is the Born cross section (first uncertainty statistical, second systematic), and $\sigma^{\mathrm{UL}}$ is the upper limit on $\sigma^{\mathrm{Born}}$ at the 90\% CL after taking into account systematic uncertainty.
}
\setlength{\tabcolsep}{1.05mm}
\renewcommand{\arraystretch}{1.2}
{
\begin{tabular}{c|cccccccccc}
\hline
$\sqrt{s}$~(MeV)& $\mathcal{L}~({\rm pb}^{-1})$ & $N_{\mathrm{obs}}$&$N_{\mathrm{sig}}$ &$N^{\mathrm{UL}}$  &$(1+\delta^{\mathrm{sig}}_{\mathrm{ISR}})$   &${(1+\delta_{\mathrm{VP}})}$  &$\epsilon~(\%)$           &$\sigma^{\mathrm{Born}}$~(pb) &$\sigma^{\mathrm{UL}}$~(pb)   \\
\hline
4843.07   &525.16 &0& $0.0^{+1.0}_{-0.0}$&$<2.0$ & 0.706  &1.056  &$10.7$&$0.0^{+1.1}_{-0.0}$ & $<2.3$ \\
4918.02  &207.82 &0&$0.0^{+1.0}_{-0.0}$  &$<2.0$ &0.750  &1.056  &$9.7$&$0.0^{+2.9}_{-0.0}$ & $<6.0$\\
4950.93   &159.28 &1&$1.0^{+1.4}_{-0.7}$&$<3.6$ &0.762  &1.056  &$9.5$ &$3.9^{+5.5}_{-2.7} \pm 0.2 $ & $<15$\\
\hline
\end{tabular}
}
\label{table:sum_result}
\end{table*}

\subsection{\boldmath $\psi_0(4360)\rightarrow \eta\psi(2S)$}
\begin{figure*}[htbp]
\centering
  \includegraphics[width=0.43\textwidth]{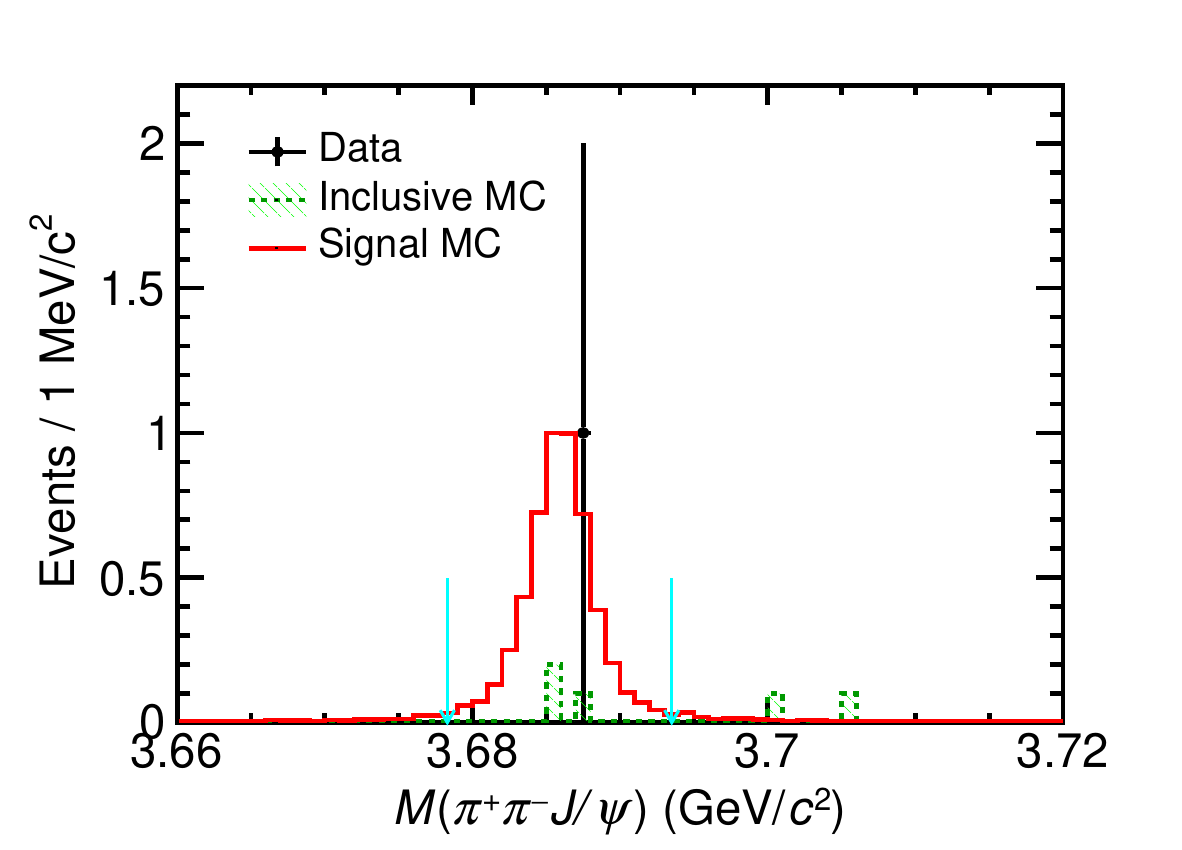}
    \includegraphics[width=0.43\textwidth]{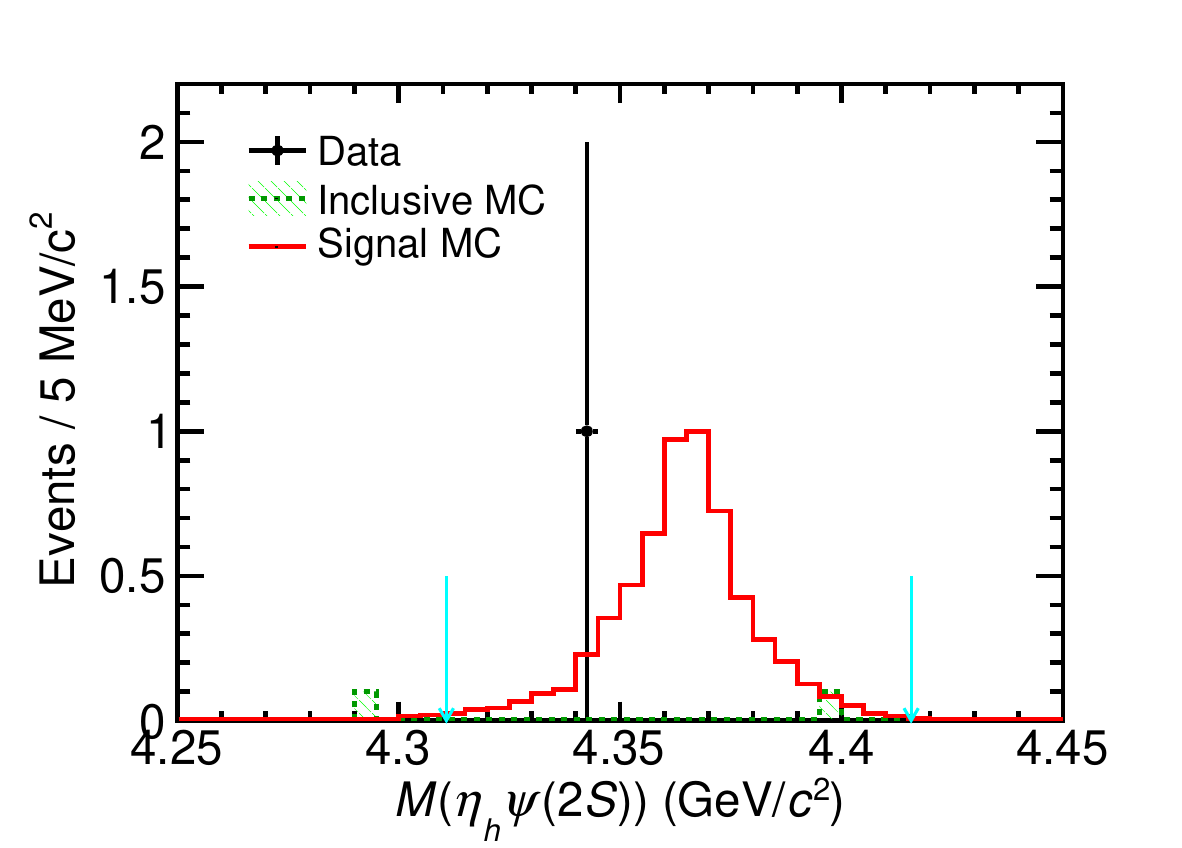}
  \caption{Distributions of $M(\pi^{+}\pi^{-}J/\psi)$ and $M(\eta_h\psi(2S))$ at the c.m. energies of 4.92 and 4.95~GeV. The black dots with error bars denote data, the green dashed histograms denote the inclusive MC and the red histograms denote the signal MC simulation of $\EE\rightarrow\eta\psi_0(4360)$. The blue solid lines mark the signal regions of $\psi(2S)$ and $\psi_0(4360)$.}
\label{fig:data_psi_2}
\end{figure*}

A search for the charmonium-like state $\psi_0(4360)$ is performed via the process $e^+e^- \rightarrow \eta \psi_0(4360)$ with $\psi_0(4360)\rightarrow \eta\psi(2S)$.
Based on MC study, the $\eta$ originating from $\psi_0(4360)$ has higher momentum than the $\eta$ produced directly from $e^+e^-$ collision. Therefore, the $\psi_0(4360)$ candidate is reconstructed from the $\eta_h \psi(2S)$ system, where $\eta_h$ denotes the higher-momentum candidate between $\eta_1$ and $\eta_2$, and its mass is defined as $M(\eta_h\psi(2S))$. 
The $\psi_0(4360)$ signal region is chosen as [4310.8, 4416.4]~MeV/$c^{2}$, corresponding to $\pm3\sigma$ around the $\psi_0(4360)$ nominal mass, where $\sigma$ is the standard deviation. 
After applying the $\psi_0(4360)$ mass window requirement, no significant signal remains.
The number of expected background events in $\psi_0(4360)$ signal region is also estimated with Eq.~(\ref{func:cal-bkg}). The numbers of background events are estimated to be 0.05 and 0.02 at the two energy points.
Figure~\ref{fig:data_psi_2} shows the $\eta_h\psi(2S)$ invariant mass distributions for data and MC samples of the processes $\EE \rightarrow \eta\psi_0(4360)$ with $\psi_0(4360)\rightarrow\eta\psi(2S)$ at c.m. energies 4.92 and 4.95 GeV. No significant $\psi_0(4360)$ candidates are found.
Therefore, the same method as previously described is employed to calculate upper limits on the numbers of signal events for these processes.  

The product of the Born cross section and branching fraction, $\sigma(e^+e^-\rightarrow\eta\psi_0(4360))\cdot\mathcal{B}(\psi_0(4360)\rightarrow\eta \psi(2S))$, at each c.m. energy is calculated using similar equation as shown in Eq.~(\ref{eq:cal_cross}). 
The ISR correction factor $(1+\delta_{\mathrm{ISR}})$ is obtained using the same method as the $\EE\rightarrow\eta\eta\psi(2S)$ process, assuming the cross section follows the lineshape of a P-wave two-body process. 
The numbers used in the calculation and corresponding upper limits at the 90\% CL
for each energy point are listed in Table~\ref{table:sum_result_two}.

\begin{table*}[htbp]
\centering
\caption{Numerical results for the process $e^+e^-\rightarrow\eta\psi_0(4360)\rightarrow\eta\eta\psi(2S)$ at each c.m. energy, where $\sqrt{s}$ is the c.m. energy, $\mathcal{L}$ is integrated luminosity, $N^{\mathrm{UL}}$ is the upper limit on the number of signal events at the 90\% CL, $(1+\delta_{\mathrm{ISR}})$ is the ISR correction factor, $(1+\delta_{\mathrm{VP}})$ is the vacuum polarization factor, $\epsilon$ is the detection efficiency, $\sigma^{\mathrm{UL}}\cdot\mathcal{B}$ is the upper limit on $\sigma(e^+e^-\rightarrow\eta\psi_0(4360))\cdot\mathcal{B}(\psi_0(4360)\rightarrow\eta \psi(2S))$ at the 90\% CL after taking into account systematic uncertainty.}
\setlength{\tabcolsep}{1.05mm}
\renewcommand{\arraystretch}{1.2}
{
\begin{tabular}{c|ccccccc}
\hline
$\sqrt{s}$~(MeV)& $\mathcal{L}~({\rm pb}^{-1})$ &$N^{\mathrm{UL}}$   &$(1+\delta^{\mathrm{sig}}_{\mathrm{ISR}})$  &${(1+\delta_{\mathrm{VP}})}$  &$\epsilon~(\%)$           &$\sigma^{\mathrm{UL}}\cdot
\mathcal{B}$~(pb)\\
\hline
4918.02  &207.82 &$<2.0$  &0.676  &1.056  &$9.6$ & $<6.8$\\
4950.93  &159.28 &$<3.6$ &0.687  &1.056  &$9.8 $ & $<16$\\
\hline
\end{tabular}
}
\label{table:sum_result_two}
\end{table*}

\section{SYSTEMATIC UNCERTAINTY ESTIMATION}

Systematic uncertainties in the cross section measurements of the processes $\EE \rightarrow \eta\eta\psi(2S)$ and $\EE \rightarrow \eta\psi_0(4360)$ with $\psi_0(4360)\rightarrow\eta\psi(2S)$ are from the luminosity measurement, the tracking of charged tracks, the photon detection, the 1C kinematic fit, the $\eta$, $J/\psi$ and $\psi(2S)$ mass windows, the $\psi_0(4360)$ mass window, the input lineshape, the generator model, the input branching fractions, and the background estimation.
The systematic uncertainties related to $\sigma(\EE\rightarrow\eta\eta\psi(2S))$ and $\sigma(\EE\rightarrow\eta\psi_0(4360))\cdot\mathcal{B}(\psi_0(4360)\rightarrow \eta\psi(2S))$ for each c.m.~energy are summarized in Table~\ref{table:sys}, where the total systematic uncertainty is calculated as a sum in quadrature of all sources of uncertainties, assuming they are independent.

\begin{table}[htbp]
\centering
\caption{Systematic uncertainties (in \%) in the cross section measurements. 
The first values in brackets are for the process $\EE\rightarrow\eta\eta\psi(2S)$, and the second for the process $e^+e^-\rightarrow\eta\psi_0(4360)\rightarrow\eta\eta\psi(2S)$. A dash indicates that a systematic effect is not applicable.}
\begin{tabular}{lccc}
\hline
Data set & 4.843 & 4.918 & 4.951\\\hline
Integrated luminosity & (0.6, -) & (0.6. 0.6) &  (0.6. 0.6) \\
Tracking & (4.0, -) & (4.0, 4.0) & (4.0, 4.0) \\
Photon detection & (1.0, -) & (1.0, 1.0) & (1.0, 1.0)  \\
1C kinematic fit & (1.0, -) & (1.1, 0.9) & (0.7, 0.9) \\
MUC requirement & (1.1, -) & (1.1, 1.1) & (1.1, 1.1) \\
$\eta_1$ mass window & (0.1, -) & (0.1, 0.1) & (0.1, 0.1) \\
$\eta_2$ mass window & (0.1, -) & (0.1, 0.1) & (0.1, 0.1) \\
$J/\psi$ mass window & (0.3, -) & (0.3, 0.3) & (0.3, 0.3) \\
$\psi(2S)$ mass window & (0.1, -) & (0.1, 0.1) & (0.1, 0.1) \\
$\psi_0(4360)$ mass window & (-, -) & (-, 5.4) & (-, 4.6) \\
ISR and generator & (2.8, -) & (1.8, 1.8) & (1.4, 1.4) \\
MC statistics & (0.9, -) & (0.9, 0.9) & (0.9, 0.9) \\
Branching fractions & (1.5, -) & (1.5, 1,5)& (1.5, 1,5) \\
Background & (0.0, -) & (0.0, 0.0) & (0.2, 0.2) \\\hline
Sum & (5.5, -) & (5.1, 7.4) & (4.9, 6.8) \\\hline
\end{tabular}
\label{table:sys}
\end{table}

The systematic uncertainties in the determination of the upper limits of the cross sections are classified as either additive or multiplicative terms. 
The additive uncertainties originate from the background estimation. 
The resulting upper limits are determined, and the most conservative upper limit is taken as the final result. 
The multiplicative systematic uncertainties are considered in the calculations of the upper limits by using {\textsc{trolke}}.

The integrated luminosity is measured using Bhabha events with an uncertainty of $0.6\%$~\cite{BESIII:2022ulv}.

The uncertainty of the pion tracking is estimated to be 1.0\% per pion based on the study of $J/\psi\rightarrow p\bar{p}\pip\pim$~\cite{BESIII:2015track_pi}.
The uncertainty of the tracking for high momentum leptons is assigned to be 1.0\% per track according to the study of $\EE\rightarrow\pip\pim J/\psi$ with $J/\psi\rightarrow\ell^+\ell^-$~\cite{BESIII:2015track_ll}. The total systematic uncertainty from the tracking of all charged tracks is assigned to be 4.0\%.

The uncertainty due to the photon detection is determined by using the control sample of $J/\psi\rightarrow\rho^0\piz$ and found to be 1.0\% per photon~\cite{photon}.
In this work, candidate events are required to have at least two photons, and only two good photon candidates are retained for further analysis.
The probability of reconstructing two photons is estimated from the photon multiplicity distribution, and the corresponding systematic uncertainty associated with photon detection is determined to be 1.0\%.

The uncertainty due to the kinematic fit is estimated by correcting the helix parameters of charged tracks according to the method described in Ref.~\cite{helix}. The difference in detection efficiencies obtained from the MC samples with and without this correction is taken as the uncertainty.

The systematic uncertainty of the MUC selection is estimated with the
control sample of $\EE \rightarrow\mu^+
\mu^-$~\cite{Zhou:2024hpq}. The difference in efficiencies between
data and MC simulation, which
is 1.1\%, is taken as the systematic uncertainty.

The uncertainties for the selection of $\eta$, $J/\psi$ and $\psi(2S)$ mass windows are estimated using the control sample of $\EE \rightarrow \piz\piz\psi(2S), \psi(2S) \rightarrow \pi^{+}\pi^{-}J/\psi$. 
The $\psi(2S)$ is reconstructed in the final state of $\pip\pim J/\psi$ with $J/\psi \rightarrow e^{+}e^{-}$ and $J/\psi \rightarrow \mu^{+}\mu^{-}$, and one $\pi^0$ is reconstructed from a photon pair, while the other $\pi^0$ is treated inclusively.
The $\piz$ signal is fitted with a signal MC shape convolved with a Gaussian function, whose parameters are left free in the fit.  
This allows the resolution difference between data and MC simulation to be determined.
The resolution difference obtained for $\pi^0$ is assumed to be representative of that for $\eta$, as both are reconstructed via similar methods. 
The MC simulated shape of $\eta$ is smeared with the Gaussian function to improve data-MC consistency. 
The resulting difference in selection efficiencies before and after smearing is taken as the systematic uncertainty associated with the $\eta$ mass window requirement.
The uncertainties for the $J/\psi$ and $\psi(2S)$ mass windows are estimated using the same method.

The uncertainty from the $\psi_0(4360)$ mass window is estimated by varying the window boundaries by $\pm1\sigma$, and the largest change in efficiency is taken as the uncertainty. 

The systematic uncertainty associated with the ISR factor and MC generator models is considered simultaneously.
The signal MC samples of the processes $e^+e^- \to \eta\eta\psi(2S)$ and $e^+e^- \to \eta\psi_0(4360)$ are generated using a three-body PHSP model and a P-wave two-body process, respectively.
Since the production threshold for $\eta \psi_0(4360)$ lies above $\sqrt{s}=4.84$ GeV, we consider two scenarios at this energy: the mass of the hypothetical charmonium-like state ($\psi_0$) is set to 4234 MeV$/c^2$ and 4295 MeV$/c^2$, corresponding to the kinematic boundaries of the $\eta\psi_0$ system. 
For the other two c.m. energy points, we take the mass of $\psi_0(4360)$ as 4366~MeV$/c^2$.
The largest difference in $(1+\delta^{\mathrm{sig}}_{\mathrm{ISR}})\times\epsilon$ between the two signal MC processes is taken as the uncertainty. 


The uncertainty due to finite MC statistics is obtained by
$\frac{1}{\sqrt{N}} \sqrt{\frac{(1-\epsilon)}{\epsilon}}$, where
$\epsilon$ is the detection efficiency and $N$ is the total number of the
generated MC events. Given that we generate 100 000 MC events at each
energy point, this corresponds to an uncertainty of approximately
0.9\%.

The branching fractions quoted from the PDG~\cite{ParticleDataGroup:2024cfk} for $\psi(2S) \rightarrow\pip\pim J/\psi$,  
$J/\psi \rightarrow e^{+}e^{-}$ and $J/\psi \rightarrow \mu^{+}\mu^{-}$ are $(34.69 \pm 0.34)$\%,
$(5.971 \pm 0.032)$\%, and $(5.961 \pm 0.033)$\%, respectively.
Since we generate signal MC events allowing both $\eta s$ to decay inclusively, the uncertainty from the branching fraction of $\eta\rightarrow\gamma\gamma$ (($39.36\pm0.18$)\%) is also included. The sum in quadrature of the individual contributions, 1.5\%, is assigned as the total systematic uncertainty due to the quoted branching fractions.

The number of expected background events is estimated according to the measured cross sections using Eq.~(\ref{func:cal-bkg}). The cross sections for the background processes are increased by their corresponding uncertainties, and the corresponding change in the signal yield is computed. The difference of the signal cross sections due to this change is taken as the uncertainty of the background estimation.

\section{SUMMARY}

In summary, the process $\EE\rightarrow\eta\eta\psi(2S)$ has been
searched for using data at the c.m.~energies of 4.84, 4.92, and
4.95~GeV. No significant signal is observed, and upper limits on
$\sigma(\EE\rightarrow\eta\eta\psi(2S))$ at the 90\% CL are
provided. The numerical results of cross sections are given in
Table~\ref{table:sum_result}. No $0^{--}$ molecular candidate
$\psi_0(4360)$ is observed at the c.m. energies of 4.92 and
4.95~GeV. Upper limits on
$\sigma(e^+e^-\rightarrow\eta\psi_0(4360))\cdot\mathcal{B}(\psi_0(4360)\rightarrow\eta
\psi(2S))$ at the 90\% CL are provided in
Table~\ref{table:sum_result_two}.  The threshold of $\eta\psi_0(4360)$
production is about 4.9 GeV, it may be highly suppressed at the
current BESIII experiment due to the limited phase-space.  At the
upcoming BEPCII-U upgrade~\cite{BESIII:2022mxl}, which has an energy
up to 5.6 GeV and has a higher luminosity than the current BEPCII, the
studies in this article can be further performed with higher energy
and larger statistics data samples.

\begin{acknowledgments}
The BESIII Collaboration thanks the staff of BEPCII (https://cstr.cn/31109.02.BEPC) and the IHEP computing center for their strong support. This work is supported in part by National Key R\&D Program of China under Contracts Nos. 2025YFA1613900, 2023YFA1606000, 2023YFA1606704; National Natural Science Foundation of China (NSFC) under Contracts Nos. 12375070, 11635010, 11935015, 11935016, 11935018, 12025502, 12035009, 12035013, 12061131003, 12192260, 12192261, 12192262, 12192263, 12192264, 12192265, 12221005, 12225509, 12235017, 12342502, 12361141819; the Chinese Academy of Sciences (CAS) Large-Scale Scientific Facility Program; the Strategic Priority Research Program of Chinese Academy of Sciences under Contract No. XDA0480600; CAS under Contract No. YSBR-101; 100 Talents Program of CAS; Shanghai Leading Talent Program of Eastern Talent Plan under Contract No. JLH5913002; Shanghai Top Talent Program of Eastern Talent Plan under Contract No. BJZH2025073; The Institute of Nuclear and Particle Physics (INPAC) and Shanghai Key Laboratory for Particle Physics and Cosmology; ERC under Contract No. 758462; German Research Foundation DFG under Contract No. FOR5327; Istituto Nazionale di Fisica Nucleare, Italy; Knut and Alice Wallenberg Foundation under Contracts Nos. 2021.0174, 2021.0299, 2023.0315; Ministry of Development of Turkey under Contract No. DPT2006K-120470; National Research Foundation of Korea under Contract No. NRF-2022R1A2C1092335; National Science and Technology fund of Mongolia; Polish National Science Centre under Contract No. 2024/53/B/ST2/00975; STFC (United Kingdom); Swedish Research Council under Contract No. 2019.04595; U. S. Department of Energy under Contract No. DE-FG02-05ER41374
\end{acknowledgments}

\end{document}